\documentclass[aaspp4]{aastex}
\usepackage{psfig,epsf}

\def\bron{XTE~J1543-568}

\shortauthors{J.J.M. in 't Zand, R.H.D. Corbet \& F.E. Marshall }

\begin{document}

\slugcomment{Accepted for publication in Astrophysical Journal Letters}

\title{DISCOVERY OF A 75~DAY ORBIT IN XTE~J1543-568}
\author{J.J.M.~in~'t~Zand\altaffilmark{1,2}}
\affil{Astronomical Institute, Utrecht University, the Netherlands}
\author{R.H.D. Corbet\altaffilmark{3,4}, F.E. Marshall\altaffilmark{5}}
\affil{Laboratory for High Energy Astrophysics,
 NASA Goddard Space Flight Center, Greenbelt, MD 20771}

\altaffiltext{1}{Also Space Research Organization Netherlands}
\altaffiltext{2}{email jeanz@sron.nl}
\altaffiltext{3}{Also Universities Space Research Association}
\altaffiltext{4}{email corbet@lheamail.gsfc.nasa.gov}
\altaffiltext{5}{email marshall@lheamail.gsfc.nasa.gov}

\begin{abstract}
Dedicated monitoring of the transient X-ray pulsar \bron\ during the first year
after its discovery has revealed the unambiguous detection of a binary
orbit. The orbital period is $75.56\pm0.25$~d, and the projected semi-major
axis 353$\pm8$ lt-sec. The mass function and position in the pulse period
versus orbital period diagram are consistent with \bron\ being a Be X-ray
binary. The eccentricity of less than 0.03 (2$\sigma$) is among the lowest
for the 12 Be X-ray binaries whose orbits have now been measured. This
confirms the suspicion that small kick velocities of neutron stars during
supernovae are more common than expected. The distance is estimated to be
larger than 10~kpc, and the luminosity at least 10$^{37}$~erg~s$^{-1}$.
\end{abstract}



\section{Introduction}
\label{secintro}

The transient X-ray source \bron\ was discovered during a slew of the
Proportional Counter Array
(PCA) on the Rossi X-ray Timing Explorer (RXTE) on Feb.~6.84, 2000,
when it had a 2 to 10 keV flux of $2\times10^{-10}$~erg~cm$^{-2}$s$^{-1}$
(Marshall, Takeshima \& In 't Zand 2000). A subsequent pointed PCA observation
on Feb.~10.69 showed a similar intensity and revealed a pulsar with
a period of 27.12$\pm0.02$~s. Later, the pulsar was found in earlier
data from the
Burst and Transient Source Experiment (BATSE) on the Compton Gamma-Ray
Observatory; the first detection of the pulsar was on Jan.~25  
(Finger \& Wilson 2000).
The most accurate location was recently determined by serendipitous
observations with the Wide Field Cameras on BeppoSAX:
$\alpha_{2000.0}~=~15^{\rm h}44^{\rm m}01^{\rm s}$,
$\delta_{2000.0}~=~-56^{\rm o}45\farcm9$ with a 99\% confidence region
delimited by a circle with a radius of 2$\farcm0$ (Kaptein, In 't Zand
\& Heise 2001). Thus far, no optical counterpart has been reported.

\bron\ continues to be active one year after discovery and is being monitored
through regular pointed PCA observations.
We here discuss the results from a timing analysis of data from the first
year of monitoring, and report the discovery of the binary orbit of \bron,
present the light curve and discuss implications for the companion of the
neutron star.

\section{Observations and data analysis}
\label{sectionobs}

A target-of-opportunity program was initiated after the discovery with RXTE to
monitor the pulsar and to try to detect Doppler delays from a binary orbit. 
The frequency of the observations went from once every day for the first 7
days, to once every 2 days during the next 10 days, to once every 3 days during
the next 75 days, and finally to once every week for the remainder of the time.
Weekly observations still continued one year after discovery. We here report
about the first year of observations (i.e., up to Feb. 3, 2001) for which
the total accumulated net exposure time is 172~ks. The number of observations
is 77. Most observations were carried out over two consecutive RXTE orbits
with a total span of 1.9~hr, but the time span has been as high as 5.5~hr.
We discuss measurements that were carried out with the PCA on RXTE.
This instrument (for a detailed description, see Jahoda et al. 1996)
consists of an array of 5 co-aligned Proportional Counter Units
(PCUs) that are sensitive to photons of energy 2 to 60 keV with a total
collecting area of 6500~cm$^2$. The spectral resolution is 18\% full-width
at half maximum (FWHM) at 6 keV and the field of view is 1$^{\rm o}$ 
FWHM. During the observations the number of active PCUs varied
between 1 and 5.

\subsection{Light curve}

We extracted spectra for each RXTE orbit of observation, for all active
PCUs combined, and for layer 1 only. All spectra were successfully modeled
between 3 and 20 keV using a power-law function with low-energy absorption
(following Morrison \& McCammon 1983), multiplied with an exponential
cutoff function and supplemented with a narrow emission line near 6.5~keV
which may be identified with Fe-K line emission for neutral to C-like
ionization states and which has a constant equivalent width of $127\pm5$~keV.
Figure~\ref{fighistory1} presents the best-fit spectral parameters and 3-20
keV flux as a function of time, with a time resolution of one observation.
The light curve exhibits variability on a time scale
of weeks. The peak was reached about 90 days after discovery and is nine
times the lowest flux levels observed thus far. Outside this main peak,
the flux occasionally reaches half the peak value. There is no obvious
periodicity.

We also studied the light curve as measured with the All-Sky Monitor on RXTE
since 1996. This light curve is quite noisy because of the small flux.
There is no conclusive evidence for outbursts other than the one reported here,
and there is no indication that \bron\ was ever brighter than the flux levels
measured with the PCA.

\subsection{Pulse period and binary orbit}

\begin{figure}[t]
  \begin{center}
    \leavevmode
\epsfxsize=13cm
\epsfbox{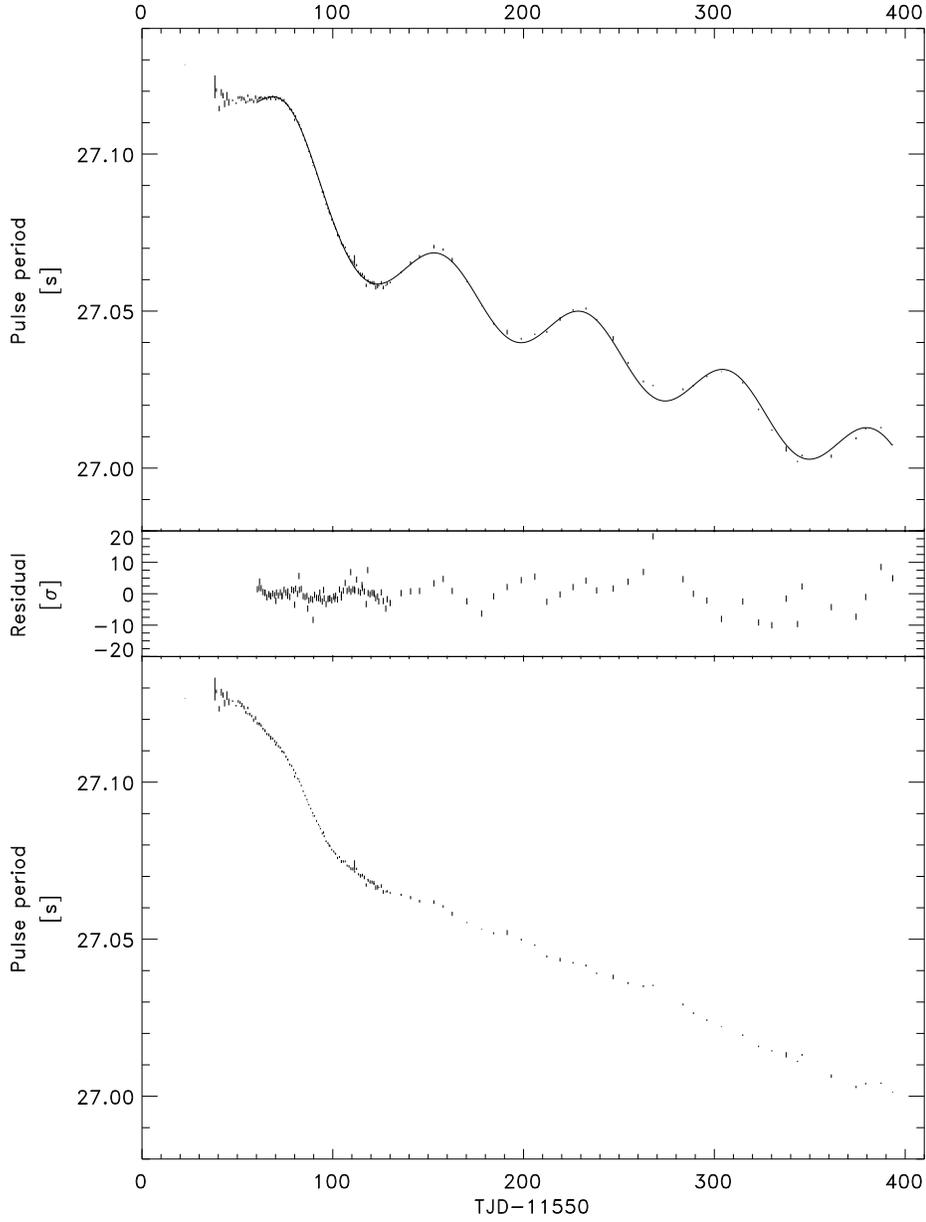}
  \caption{Time history of pulse period as measured with the PCA and BATSE.
The solid curve shows
the fitted model (see text) that was applied to data after TJD~11610.
The middle panel shows the residuals of the data points with respect
to the fitted model in terms of $\sigma$. The bottom panel shows the
pulse period history after subtraction of the sinusoidal variation.
All error bars are without multiplication by 3.2 (see text).
\label{fighistory2}
}
  \end{center}
\end{figure}

For the study of the pulse period, we employed event data with 0.5~s time
resolution and 64 readout channels for photon energy (i.e., configuration
{\tt E\_500us\_64M\_0\_1s}). We optimized the accuracy of the
pulse period determination as a function of detector layers and bandpasses
and find the best result if all layers are included and photons with
energies higher than 20 keV excluded, although this optimum is only marginally
better than if only layer 1 is included. We combined the data from all 
active PCUs, applied standard filters to
eliminate bad data, and corrected the resulting light curves to the
solar system barycenter. The latter includes a correction for the satellite
orbit. For each of the observations, we measured the pulse period by
epoch folding the data for a relevant range of test periods and calculating
the $\chi^2$ values for a constant flux hypothesis. The resulting curves 
show sinc-like trends.
To obtain the best-fit period we fitted for each observation the peak curve
with a 3rd degree polynomial and identified the period with the peak position
of the cubic function. The 68\% confidence error was estimated by determining
the range of periods for which the $\chi^2$ value is within 1 from the
maximum. This procedure was verified through Monte Carlo simulations.
The resulting measurements are presented in figure~\ref{fighistory2}. They are
supplemented with 78 BATSE measurements from the public domain\footnote{
made available through the WWW at URL {\tt 
http://www.batse.msfc.nasa.gov/batse/}} and from Finger \& Wilson (2000),
pertaining to TJD~11572-11678 (TJD~=~JD$-2,440,000.5$).
The BATSE observations stopped prior to de-orbiting the satellite in June 2000.
Figure~\ref{fighistory2} clearly shows that there is a 75~d periodicity which
must be due to an expected binary orbit. Also, long time scale spin-up trends
are visible which are expected to result from accretion torque. 
The spin-up is faster in the first $\sim100$~days which
is consistent with an accretion torque explanation because the flux
and, therefore, the accretion rate is larger at that time.

\begin{figure}[t]
  \begin{center}
    \leavevmode
\epsfxsize=13cm
\epsfbox{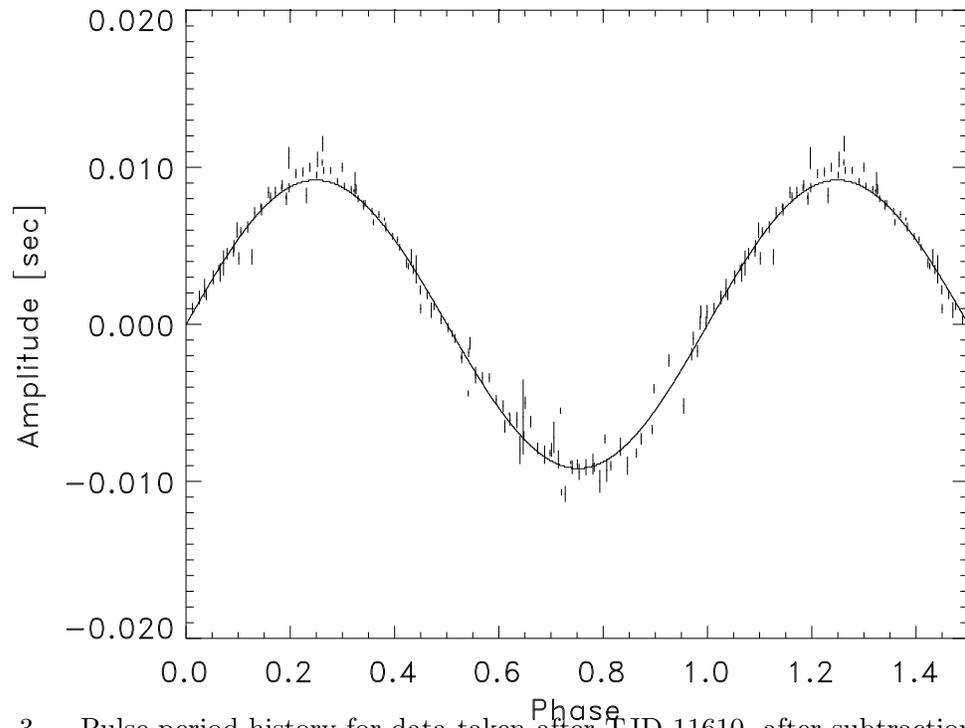}
  \caption{Pulse period history for data taken after TJD~11610, after
subtraction of spin-up model and folding with period of 75.56 days.
The solid curve shows the solution for the binary orbit, with $e=0.005$.
All error bars are without multiplication by 3.2 (see text).
\label{figfold}
}
  \end{center}
\end{figure}

In order to disentangle the orbital modulation from the
spin up trend, we modeled in an initial step
the orbital modulation through a simple sinusoidal variation. The spin
up was modeled through various simple functions with similar success and we
here present the results when the period derivative $\dot{P}$ is described
by a Gaussian function on top of a flat, non-zero, level. We only modeled
the data after TJD~11610 because the spin up seems to be more complex
before that. For the determination of the orbit this is not detrimental
because the excluded time interval represents only about 5\% of the complete
time span of the observations. The results of the modeling are presented in
figure~\ref{fighistory2}. In a second step, we subtracted the model for the
spin up from the data and fitted a model of an eccentric binary orbit to
the residuals. The best-fit $\chi^2_\nu$ is rather high at 10.2 (for 118
degrees of freedom $\nu$), which we attribute to an incomplete spin-up model.
In order to obtain estimates for the errors in the
model parameters that take into account the systematic uncertainties, 
we forced $\chi^2_\nu$ to 1 through a multiplication of
all pulse period errors with a constant factor $\sqrt{10.2}=3.2$. The results
for the orbital and spin-up parameters are given in table~\ref{pulsefit}. 
In figure~\ref{fighistory2} (bottom panel) we plot the data after subtraction
of the orbital modulation model, in figure~\ref{figfold} after subtraction of
the spin-up model and folding with the orbital period. As a double check,
we carried out the same procedure for the data after TJD~11670 which
excludes the times when the Gaussian component of the spin-up model
is important. Thus, possible strong systematic errors are avoided at the
expense of accuracy. The result, given in an extra column in
table~\ref{pulsefit}, is consistent with the previous result.

\begin{table}[]
\caption[]{Fit parameters to pulse period history\label{pulsefit}. The errors
are for 68\% confidence. The numbers in the third column are, as a reference,
from data after TJD~11670 (40 data points)}
\begin{tabular}{lll}
\hline\hline
Orbital period $P_{\rm orb}$ (days) &75.56$\pm0.25$  & 75.88$\pm0.40$\\
Orbital amplitude (ms)& 9.18$\pm0.21$ & 9.07$\pm0.26$\\
Semi-major axis $a_{\rm x~sin}i$ (lt-sec)& 353$\pm8$ & 349$\pm20$ \\
Eccentricity $e$ & $<0.03$ at 2$\sigma$ &$<0.03$ at 2$\sigma$\\
Epoch of mean longitude $\pi/2$ (TJD)& 11725.88$\pm0.16$ & 11725.54$\pm0.62$\\
Pulse period $P$ at TJD 11600 (s) &27.12156$\pm0.00059$ & 27.1225$\pm0.0043$\\
Constant $\dot{P}$ term (s~s$^{-1}$)& $-(2.84\pm0.02)\times10^{-9}$ &$-(2.84\pm0.03)\times10^{-9}$ \\
Gaussian $\dot{P}$  term & &\\
\hspace{0.5cm}$\sigma$ (days) & 13.30$\pm$0.02&\\
\hspace{0.5cm}peak time (TJD)& 11636.71$\pm$0.02&\\
\hspace{0.5cm}peak $\dot{P}$ value (s~s$^{-1}$)& $-(1.27\pm0.03)\times10^{-8}$  &\\
\hline\hline
\end{tabular}
\end{table}

\begin{figure}[t]
  \begin{center}
    \leavevmode
\epsfxsize=13cm
\epsfbox{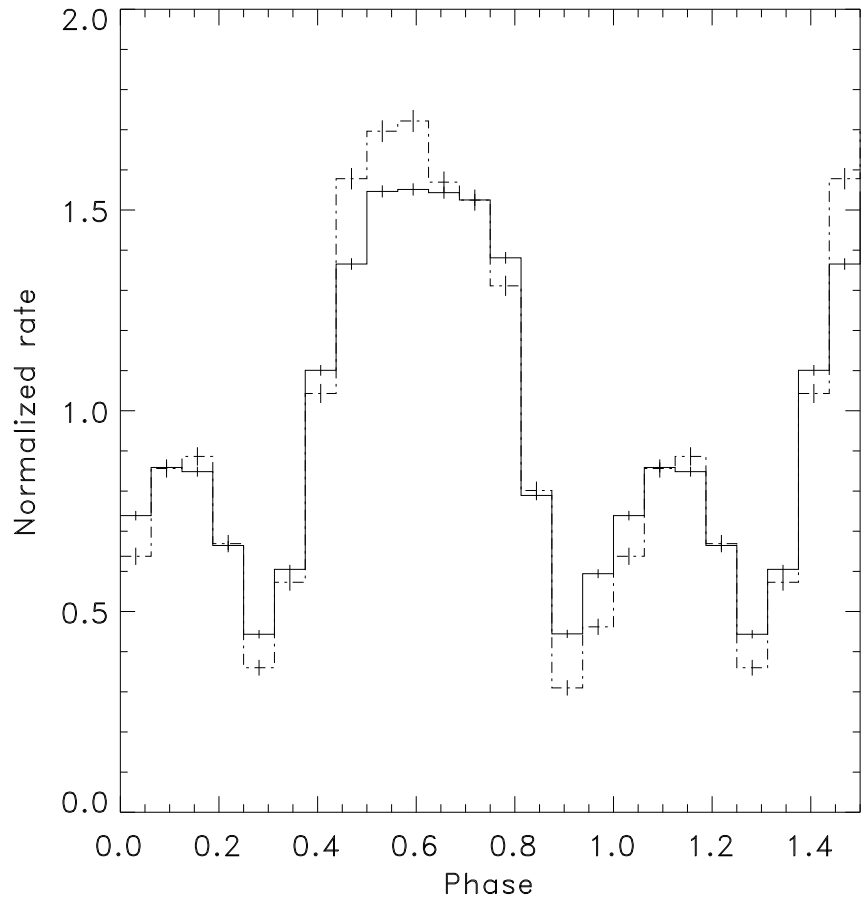}
  \caption{Pulse profiles for the observation taken on TJD 11637, when
the peak flux was reached. The solid histogram is for 3-10 keV, the
dashed one for 10-20 keV.
\label{figprofile}
}
  \end{center}
\end{figure}

The pulse profile was studied for 3 observations throughout the total
observation span, see figure~\ref{figprofile} for the profile at flux peak. The
profile consists of two broad peaks at a phase difference
of 0.5 and with different magnitudes. The pulsed fraction is
consistently 65$\pm5$\%. This kind of profile and amplitude
is not unusual for pulsars among high-mass X-ray binaries (c.f., Bildsten
et al. 1997 and White et al. 1995). We determined the profiles in
two energy bands (2 to 10 keV and 10 to 20 keV) and
do not find strong morphological differences. The pulse fraction is
about 10\% higher in the upper energy band.

\section{Discussion}
\label{sectionana}

Given the projected semi-major axis $a_{\rm x} {\rm sin}i$ and $P_{\rm orb}$,
we can place constraints on the companion of the neutron star. The mass
function is
\begin{eqnarray}
f(M) & = & \frac{4\pi^2(a_{\rm x}{\rm sin}i)^3}{GP^2_{\rm orb}} =
\frac{(M_{\rm c} {\rm sin}i)^3}{(M_{\rm x} + M_{\rm c})^2} = 8.2\pm0.5~{\rm M_\odot}
\end{eqnarray}
If $M_{\rm x}=1.4$~M$_\odot$, this implies a minimum mass for
the companion star of 10.0~M$_\odot$. For a main sequence
star, the implied spectral type is B3 or earlier (Zombeck 1990).
The orbital period places \bron\ in the 'Corbet' diagram of pulse period
versus orbital period unambiguously on the branch of optically confirmed Be 
transients (Corbet et al. 1986, c.f. Bildsten et al. 1997).

\begin{figure}[t]
  \begin{center}
    \leavevmode
\epsfxsize=13cm
\epsfbox{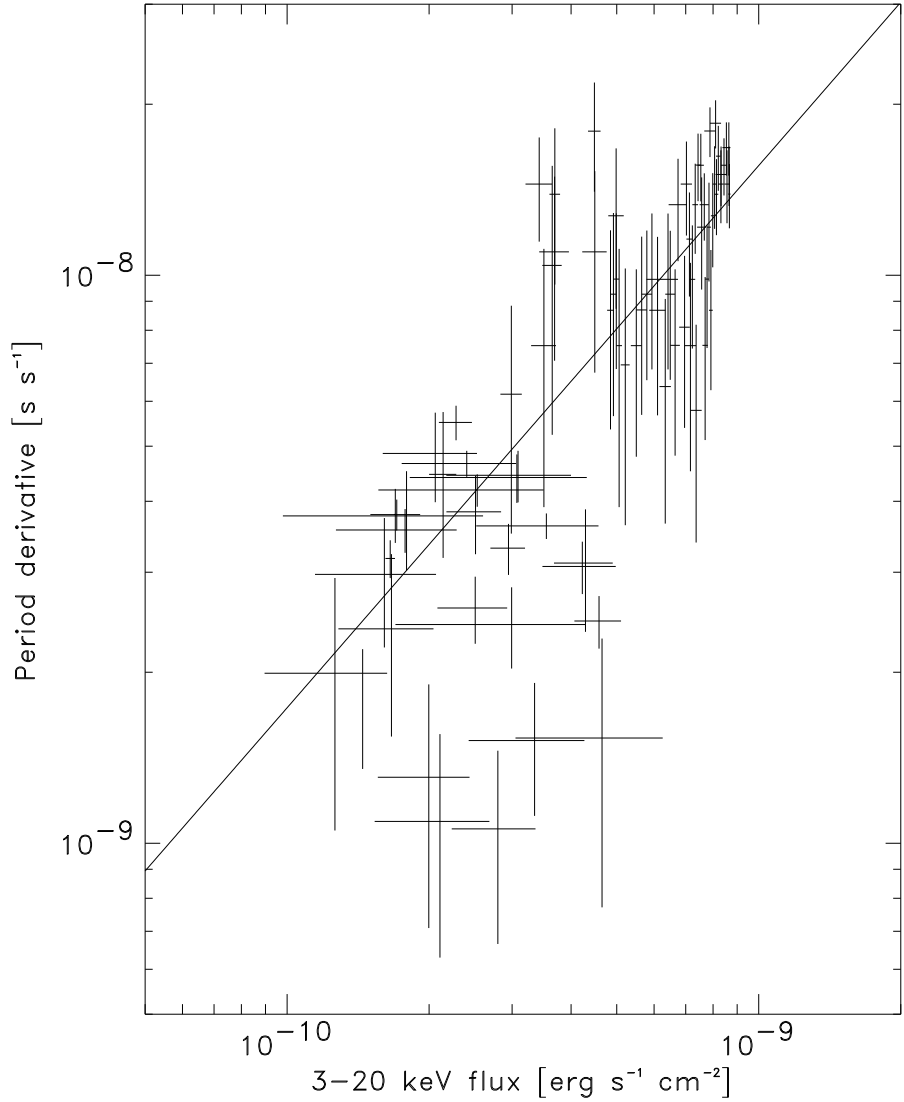}
  \caption{Period derivative, as determined from BATSE and RXTE data
points, versus 3-20 keV flux. The baseline for each period derivative
determination was three observations (or at least 2 days). The error bars
on the flux represent the range of fluxes measured during this baseline.
The solid line is the best fit power-law function, with index
$0.95\pm0.06$.
\label{figghosh}
}
  \end{center}
\end{figure}

In principle the distance is constrained by the spin-up rate and flux.
The flux is a measure of the mass accretion rate which
determines the torque on the neutron star and the spin-up rate.
Given the steadiness of the spin-up rate, we consider it likely that the
accretion occurs through a disk.  In that case Ghosh \& Lamb (1979)
predict a spin-up rate of
\begin{eqnarray}
-\dot{P} & = & 1.9\times10^{-12} \mu_{30}^{2/7} m^{-3/7} R_6^{6/7} I_{45}^{-1}
P^2 L_{37}^{6/7} \hspace{1cm} {\rm s~s}^{-1}
\end{eqnarray}
for a neutron star with magnetic dipole moment $\mu_{30}$ in units of
10$^{30}$~G~cm$^3$, mass $m$ in units of 1.4~M$_\odot$, radius $R_6$ in units
of 10~km, moment of inertia $I_{45}$ in units of 10$^{45}$~g~cm$^2$, and
luminosity $L_{37}$ in units of 10$^{37}$~erg~s$^{-1}$. In
figure~\ref{figghosh} we present the period
derivative as a function of 3-20 keV flux. Our data are rather noisy
and do not cover a large dynamic range in flux, but they
appear to be roughly consistent with a 6/7 power law relationship.
Given the spectral shapes found,
the overall bolometric correction is about a factor of 2. If we calibrate
the above $-\dot{P}/L_{37}$ relation at the $-\dot{P}=1\times10^{-8}$
s~s$^{-1}$ level and assume all normalized parameters to be 1,
the implied distance is 26~kpc. This would place \bron\ at an unlikely
position. Acceptable distances would perhaps be at least about 50\% smaller.
It is hard to achieve that by adjusting the other parameters individually:
$\mu_{30}$ would have to be at least 64 times larger,
$m$ 16 times smaller, $R_6$ 4 times larger,
or $I_{45}$ 3 times smaller. Therefore, given the correctness
of the Ghosh \& Lamb formulation, there is at least the suggestion that the
distance is large.

Another constraint for the distance, that does not require
$\mu$, may be obtained from an argument over
the magnetospheric radius $r_{\rm mag}$ at which the neutron star
magnetosphere begins to dictate the mass flow. This radius scales
with $\dot{M}^{-2/7}$ or $F^{2/7}$. 
If the flux $F$ varies between $F_{\rm min}$ and
$F_{\rm max}$ at times when the mass accretion persists, 
$r_{\rm mag}<r_{\rm co} (F_{\rm min}/F_{\rm max})^{2/7}$ at peak flux, where
$r_{\rm co}$ is the co-rotation radius at which the
Keplerian period is equal to the spin period
($r_{\rm co}=(GMP^2/4\pi^2)^{1/3}$).
The torque $2 \pi I \dot{\nu}$ is given by $\dot{M}\sqrt{GMr_{\rm mag}}$.
Substition in the above condition reveals
\begin{eqnarray}
\dot{M} & > & - 5.6\times10^2 \frac{I_{45}\dot{P}}
                 {m^{2/3} P^{7/3} (F_{\rm min}/F_{\rm max})^{1/7}} 
                 \hspace{0.5cm} {\rm M}_\odot~{\rm yr}^{-1}
\end{eqnarray}
For standard parameters and $F_{\rm max}=4.5F_{\rm min}$ (after TJD~11670),
we obtain $\dot{M}>9.8\times10^{-10}$~M$_\odot$yr$^{-1}$. If all liberated
potential energy is released in the form of radiation, the bolometric
luminosity is constrained by $L>1.2\times10^{37}$~erg~s$^{-1}$.
For $F_{\rm max}=4.5\times10^{-10}$~erg~s$^{-1}$cm$^{-2}$ and a
bolometric correction of 2, a minimum distance of 10~kpc is implied.
Again this is a fairly large distance.

\begin{figure}[t]
  \begin{center}
    \leavevmode
\epsfxsize=13cm
\epsfbox{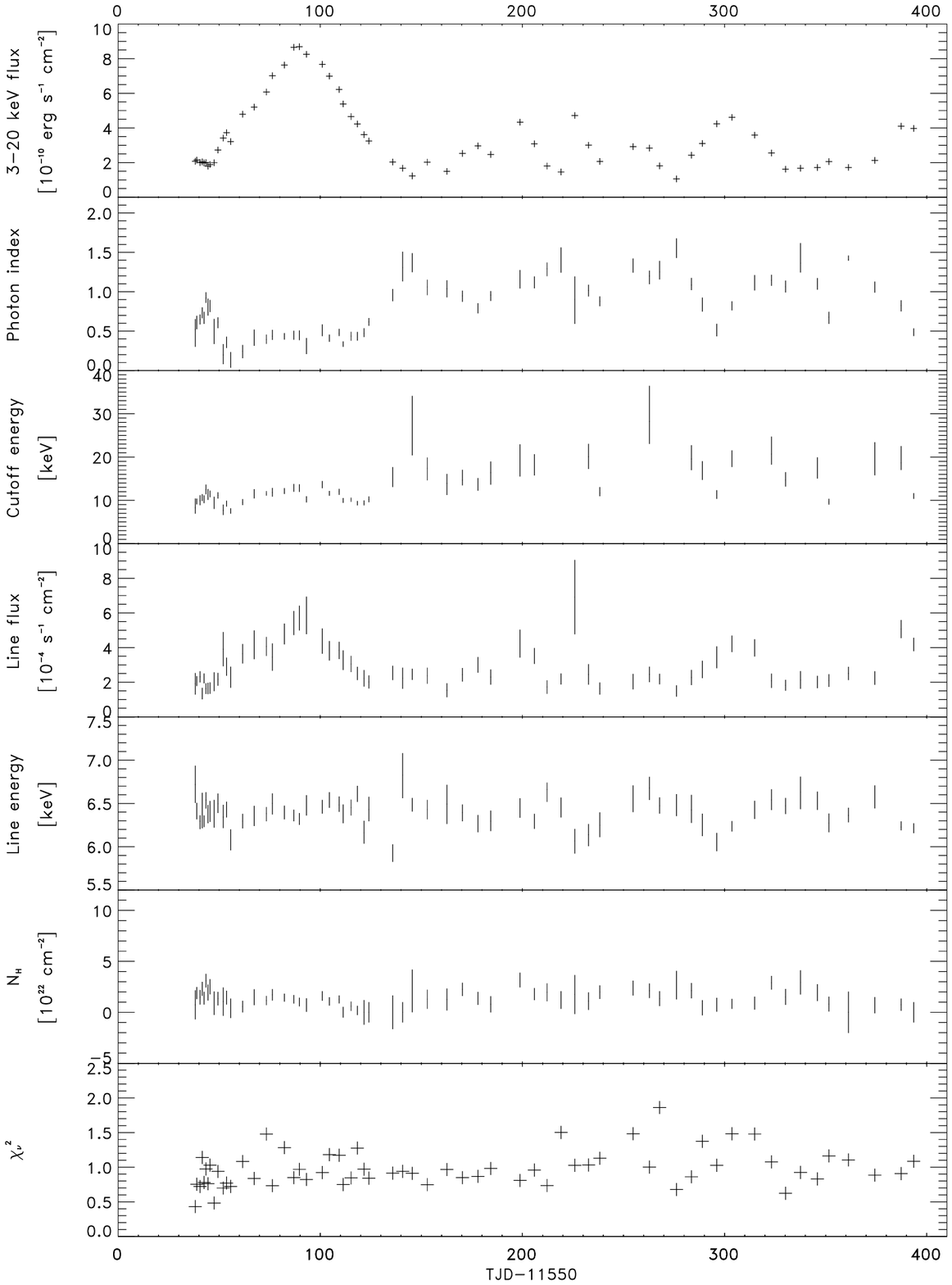}
  \caption{Top panel: PCA-measured time history of \bron\ flux in 3 to 20 keV,
from the time of discovery through the first year of monitoring
observation. The errors (not indicated) are typically 1\%.
Second through sixth panel: evolution of power-law photon index, e-folding
cutoff energy, photon flux
and centroid energy of the emission line, and hydrogen column density. For
some spectra, the cutoff energy could not be constrained
and no data point is given in the appropriate panel.
The average hydrogen column density is
$(1.3\pm0.1)\times10^{22}$~cm$^2$. The emission line flux follows the
overall flux. In fact, the equivalent width is consistent with
being constant with a value of 127~eV. Bottom panel:
$\chi^2_\nu$ statistic for the relevant spectral fits.
\label{fighistory1}
}
  \end{center}
\end{figure}

These distance estimates suggest that
searches for the optical counterpart may prove difficult, also
because the large value of $N_{\rm H}$ points to a high extinction.
For a 10.0~M$_\odot$ main sequence star, $M_{\rm V}=-2$. If
$N_{\rm H}=1.3\times10^{22}$ (see figure~\ref{fighistory1}; this is
consistent with the total Galactic column as inferred from maps
by Dickey \& Lockman 1990),
$A_{\rm V}=7.4$ (according to the standard conversions introduced by
Predehl \& Schmitt 1995). For a distance of 26~kpc, the apparent
visual magnitude would be $V=23$, for a distance of 10~kpc, $V=21$.

The most interesting result of our analysis is the low eccentricity.
It is among the lowest for the 12 confirmed or
suspected Be X-ray binaries whose orbits have now been determined
(8 cases are reviewed by Bildsten et al. 1997 and since then 
orbits have been determined for 2S 1845-024 by Finger et al. 1999,
for SAX~J2103.5+4545 by Baykal, Stark \& Swank 2000, and for X Per by
Delgado-Mart\'{\i} et al. 2001). There is one other system apart from
\bron\ which has an eccentricity that is consistent with 0 ($e<0.09$ at
$2\sigma$ confidence for 2S~1553-54, see Kelley et al. 1983). However,
one should note that Kelley et al. have a sparse coverage of less than
one binary orbit
and do not solve for spin-up. Finding low-eccentricity
systems is remarkable because one would expect most long-period
Be systems to have undergone supernovae with large kick velocities for
the neutron stars and to have circularization time scales far longer than
their age. \bron\ confirms the suspicion that there may be a substantial
population of neutron stars formed with little or no kick (Delgado-Mart\'{\i}\
et al. 2001).


\acknowledgements

We thank the project scientist Jean Swank for continuing the RXTE
observations of \bron\ through
public TOOs after our TOO program time was spent, Evan Smith for his tireless
efforts to schedule the many observations, Mike Tripicco for his
assistance in the barycentric corrections for 2001 PCA data, Ron
Remillard for his advice on the ASM light curve, and the referee Mark Finger
for valuable suggestions. JZ acknowledges financial support
from the Netherlands Organization for Scientific Research (NWO).



\end{document}